\documentclass[preprint,showpacs,preprintnumbers,prb,fleqn,floatfix,superscriptaddress,longbibliography]{revtex4}
\usepackage{graphicx}
\usepackage{amssymb}
\usepackage{epsfig}
\usepackage{amsmath}
\usepackage{bm}
\usepackage[english]{babel}

\usepackage[breaklinks]{hyperref}
\usepackage[all]{hypcap}

\hypersetup{
    plainpages=false,
    bookmarks=false,         
    unicode=false,          
    pdftoolbar=true,        
    pdfmenubar=true,        
    pdffitwindow=false,     
    pdfstartview={FitH},    
    pdfnewwindow=true,      
    linktoc=section,
    colorlinks=true,       
    linkcolor=blue,          
    citecolor=blue,        
    filecolor=magenta,      
    urlcolor=blue           
}

\begin{document}


\title{Spatially resolved studies of the phases and morphology of methylammonium and formamidinium lead tri-halide perovskites}

\author{K. \surname{Galkowski}}
\affiliation{Laboratoire National des Champs Magn\'etiques Intenses,
CNRS-UJF-UPS-INSA, 143, avenue de Rangueil, 31400
Toulouse}\affiliation{Institute of Experimental Physics, Faculty of
Physics, University of Warsaw - Pasteura 5, 02-093 Warsaw, Poland}

\author{A.  \surname{Mitioglu}}
\affiliation{Laboratoire National des Champs Magn\'etiques Intenses,
CNRS-UJF-UPS-INSA, 143, avenue de Rangueil, 31400 Toulouse}
\affiliation{Institute of Applied Physics, Academiei Str. 5,
Chisinau, MD-2028, Republic of Moldova}

\author{A.  \surname{Surrente}}
\affiliation{Laboratoire National des Champs Magn\'etiques Intenses,
CNRS-UJF-UPS-INSA, 143, avenue de Rangueil, 31400 Toulouse}

\author{Z.  \surname{Yang}}
\affiliation{Laboratoire National des Champs Magn\'etiques Intenses,
CNRS-UJF-UPS-INSA, 143, avenue de Rangueil, 31400 Toulouse}

\author{D. K. \surname{Maude}}
 \affiliation{Laboratoire National des Champs Magn\'etiques Intenses,
CNRS-UJF-UPS-INSA, 143, avenue de Rangueil, 31400 Toulouse}

\author{P. \surname{Kossacki}}
\affiliation{Institute of Experimental Physics, Faculty of Physics,
University of Warsaw - Pasteura 5, 02-093 Warsaw, Poland}

\author{G. E. \surname{Eperon}}
 \affiliation{University of Oxford, Clarendon
Laboratory, Parks Road, Oxford, OX1 3PU, United Kingdom}

\author{J. T-W.\surname{Wang}}
 \affiliation{University of Oxford, Clarendon
Laboratory, Parks Road, Oxford, OX1 3PU, United Kingdom}

\author{H. J. \surname{Snaith}}
\affiliation{University of Oxford, Clarendon Laboratory, Parks Road,
Oxford, OX1 3PU, United Kingdom}

\author{P. \surname{Plochocka}}
\email{paulina.plochocka@lncmi.cnrs.fr}\affiliation{Laboratoire National des Champs Magn\'etiques Intenses,
CNRS-UJF-UPS-INSA, 143, avenue de Rangueil, 31400 Toulouse}

\author{R. J. \surname{Nicholas}}
\email{robin.nicholas@physics.ox.ac.uk}\affiliation{University of
Oxford, Clarendon Laboratory, Parks Road, Oxford, OX1 3PU, United
Kingdom}

\date{\today }

\begin{abstract}
The family of organic-inorganic tri-halide perovskites including MA (MethylAmmonium)PbI$_{3}$, MAPbI$_{3-x}$Cl$_{x}$, FA (FormAmidinium)PbI$_{3}$ and FAPbBr$_{3}$ are having a tremendous impact on the field of photovoltaic cells due to their ease of deposition and efficiencies, but device performance can be significanly affected by inhomogeneities. Here we report a study of temperature dependent micro-photoluminescence which shows a strong spatial inhomogeneity related to the presence of microcrystalline grains, which can be both light and dark. In all of the tri-iodide based materials there is evidence that the tetragonal to orthorhombic phase transition observed
around 160K does not occur uniformly across the sample with domain formation related to the underlying microcrystallite grains, some of which remain in the high temperature, tetragonal, phase even at very low temperatures. At low temperature the tetragonal domains can be significantly influenced by local defects in the layers. In FAPbBr$_{3}$ a more macroscopic domain structure is observed with large numbers of grains forming phase correlated regions.
\end{abstract}

\maketitle

Organic inorganic perovskite based solar cells have recently reached power conversion efficiencies of over 20$\%$
\cite{Zhou14,Graetzel14}. The simple and cheap fabrication of the devices on non-crystalline substrates by vapor
deposition \cite{Liu13} or solution processing\cite{Lee12,Burschka13} together with the tunability of the
bandgap by chemical substitution~\cite{Noh13,Eperon14a,xing14} has made organic inorganic perovskites very attractive
materials not only for photovoltaic applications (PV), but also in light emitting diodes\cite{Tan14},
lasers\cite{Deschler14,Saliba15} and photodetectors\cite{Stranks15}. An immense effort has been made to investigate both
the electronic and structural properties of these materials which has demonstrated several of their remarkable characteristics which makes them ideally suited for device applications such as, their long diffusion
lengths\cite{Xing13,Stranks13}, the small binding energies of the excitons \cite{Galkowski16,Miyata15} and radiative life times for the carriers exceeding 500 ns \cite{Stranks13,Zhou14,deQuilettes15,Marco16}. Improving the
performance of devices is leading many groups now to focus on understanding the influence of the growth conditions
\cite{Eperon14,Burschka13,Zhou14} and passivation of the surface to suppress non radiative trap/recombination of the
carriers~\cite{Abate14,Noel14,Chen14a,Marco16}.

However, to date relatively few studies have been performed to elucidate the
influence of the microscopic structure of the perovskites on their
electronic properties and crucially the performance of
devices. Only recently, confocal microscopy correlated with
scanning electron microscopy (SEM) has been employed to
investigate the microscopic structure of
these materials\cite{deQuilettes15}. It was shown that the emission
intensity and life time of the carriers vary with position and are correlated to the presence of microcrystalline grains at a scale of a few micrometers. The weaker emission is correlated with a reduced carrier lifetime which enabled the presence of specific dark grains to be identified. In a similar microscopic
study, an improvement of the life time of the carriers by
employing guanidinium to reduce the number and influence of the dark grains was shown to significantly improve device performance~\cite{Marco16}. These results
show the importance of the morphology of the material, since the
presence of the dark grains, acting as non radiative traps,
seriously degrades the performance of devices.

A further complication in understanding the properties of the
organic-inorganic tri-halide perovskites is that they show phase transitions at lower temperatures
 to structures with reduced symmetry. The tri-iodides transform from a
 tetragonal to an orthorhombic structure at around 150 K accompanied by
 an abrupt increase in band gap\cite{Baikie13,Incenzo14,Even14,Yamada15} of $\sim$ 100 meV.  The pure tri-bromides
  are cubic at room temperature\cite{Eperon14, Noh13}, and transform to tetragonal
  in the alloy family FAPbBr$_{3-x}$I$_{x}$ at around $x$=0.6\cite{Eperon14}.
  As a function of temperature the pure tri-bromides become tetragonal at around 240 K and then orthorhombic below
  150 K\cite{Onada90}. The transition from the cubic to the tetragonal structure
  appears to have no significant influence on the band gap\cite{Kunugita15, Galkowski16},
  but the transition to the orthorhombic structure only
  produces a small increase of around 10 meV.


The temperature dependence of the band gap has previously been
studied by absorption in large area samples, where the phase transition causes a continuous change of the energy of a single
edge~\cite{Incenzo14,Yamada15,Galkowski16}, which is most abrupt
for MAPbI$_{3}$ and FAPbBr$_{3}$ while for the mixed halide alloy
MAPbI$_{3-x}$Cl$_{x}$ the shift of the absorption energy is quite
extended over a temperature range of the order of 50 K\cite{Galkowski16}. In the case of
MAPbI$_{3}$, it has been possible to observe the
presence of two absorption edges within a few degrees of the phase
transition as shown in the SI here and reported previously\cite{Yamada15}. This has been attributed to a coexistence
of the tetragonal and orthorhombic phases for a small temperature range close to the phase transition. In MAPbI$_{3-x}$Cl$_{x}$, however, temperature and intensity dependent photoluminescence (PL) measurements on large area samples have been used to infer the presence of microcrystalline inclusions of the high temperature tetragonal phase at temperatures down to 4\,K \cite{Wehrenfennig14b}.

In this paper, we report a detailed microscopic study of the
family of organic-inorganic tri-halide perovskites. The chemical
structure of the samples is ABX$_{3}$ where
A=CH$_{3}$NH$^{+}_{3}$ = MA (MethylAmmonium) or
A=CH(NH$_{2}$)$_{2}$ = FA (FormAmidinium), B=Pb$^{2+}$; and X =
Cl$^{-}$, I$^{-}$ or Br$^{-}$, or an alloyed combination of
these. We have employed spatially resolved micro
photoluminescence ($\mu$PL) to locally probe the optical response
of the films on a micrometer scale in a range of temperatures from
4 K up to 300 K. All the samples studied show the presence of both
dark and bright grains, independently of the temperature.
Additionally, in all of the compounds investigated there is evidence
that the tetrahedral to orthorhombic phase transition observed
around 160 K does not occur uniformly across the sample. Domains of
the high temperature phase remain even at very low temperatures and the relative importance and behaviour of these can be significantly altered by the presence of local perturbations to the layer morphology.

\begin{figure}[h]
\includegraphics[width=15cm]{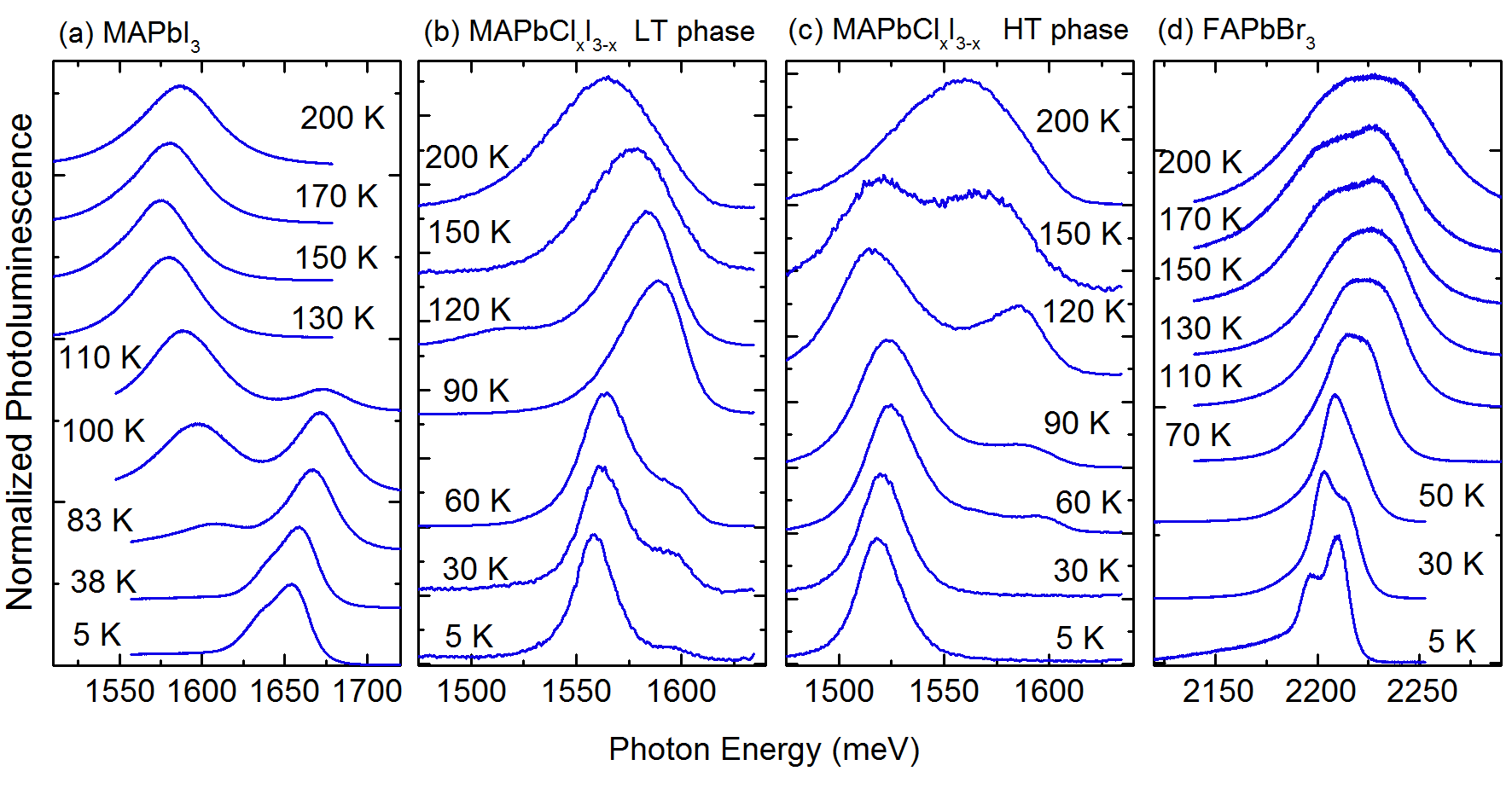}
\caption{\label{fig1} \textbf{Temperature evolution of the micro
photoluminescence for selected perovkites}: Temperature
dependence of the $\mu$PL spectra for (a) MAPbI$_{3}$,
(b) and (c) two different spot positions on MAPbI$_{3-x}$Cl$_{x}$ corresponding to domains of the orthorhomic and tetragonal phases and (d) FAPbBr$_{3}$.}
\end{figure}

\begin{figure}[h]
\includegraphics[width=15cm]{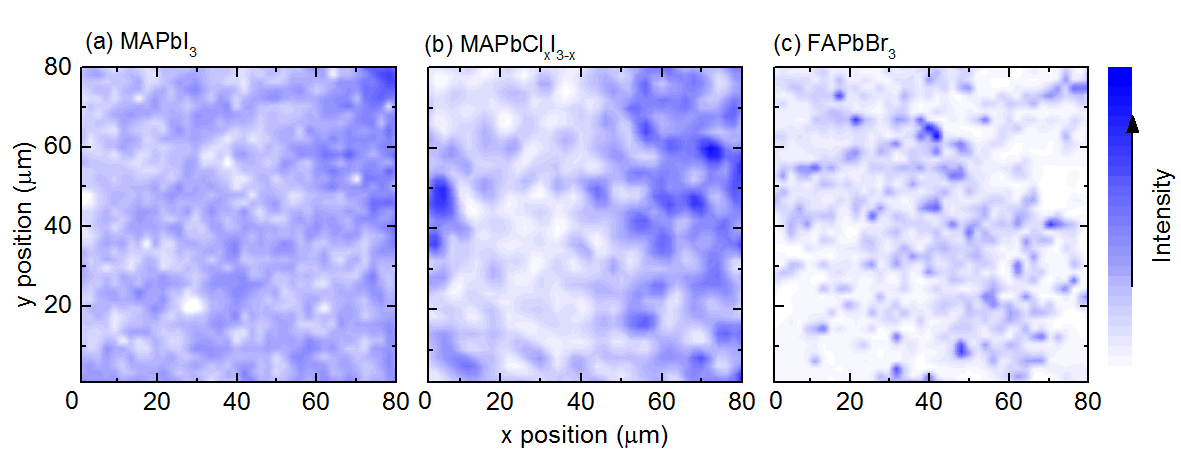}
\caption{\label{fig2} \textbf{Room temperature
micro-photoluminescence maps for selected perovkites}:
Integrated intensity of the $\mu$PL peak as a function of the position for (a) MAPbI$_{3}$, spot illumination intensity 40 nW, (b) MAPbI$_{3-x}$Cl$_{x}$ , spot illumination intensity 60 nW, and (c) FAPbBr$_{3}$, , spot illumination intensity 100 nW,
respectively at 300K. The scanning step is $2 \mu$m}
\end{figure}

The $\mu$PL reveals that the emission behaviour is strongly dependent on both temperature and position. At high
temperatures($>$200\,K), the spectra show only a single emission peak at an
energy $\sim$ 20 meV below the band gap deduced from absorption as shown in Fig\ref{fig1}, suggesting
that emission is dominated by an essentially free exciton state.
The room temperature integrated emission maps are shown in Fig.~\ref{fig2} (a) - (c)
for MAPbI$_{3}$, MAPbI$_{3-x}$Cl$_{x}$ and FAPbBr$_{3}$
respectively, where the intensity varies across the scanned area
showing brighter and darker emission areas. This is similar to previous
observations of dark and bright grains for
MAPbI$_{3}$\cite{deQuilettes15}. All three compounds show similar
behaviour, consistent with the microcrystalline grain structure known to be present
in such thin layer perovskites, however it is noticeable that the microcrystallites for the mixed halide MAPbI$_{3-x}$Cl$_{x}$ are significantly larger, consistent with its usually better device performance and photoluminescence emission\cite{Stranks15, Stranks13}.

\begin{figure}[h]
\includegraphics[width=15cm]{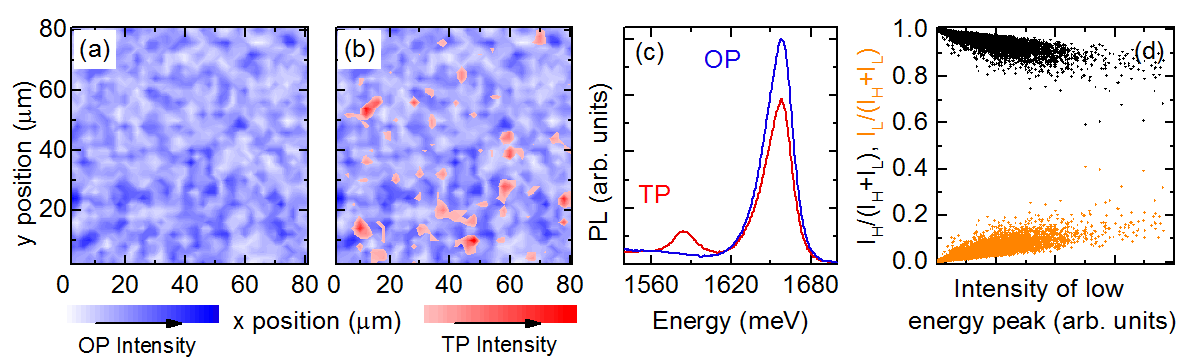}
\caption{\label{fig3} \textbf{Spatially resolved
micro-photoluminescence for MAPbI$_{3}$}:(a) Integrated intensity of
the high energy (OP) peak as a function of position. The intensity increases
from white to dark blue. The scanning step is $2 \mu$m. (b) A second map of the low energy (TP) peak is superimposed onto the map shown in (a). The intensity increases from white to red. (c) Typical spectra for positions where
only the OP peak was observed (blue) and where both peaks were observed
(red). (d) Correlation of the integrated intensity of the TP
(orange) and OP (black) peaks versus intensity of the low energy TP peak. All the measurements have been performed at 4K, spot illumination intensity 5 nW. }
\end{figure}

At lower temperatures there are significant changes in behaviour as
the spectral emission develops considerable positional dependence
below the tetragonal to orthorhombic phase transition temperature of 155K.  Fig.~\ref{fig1}(a) shows spectra at a typical location for MAPbI$_{3}$ as a function of temperature. At around and below the temperature where the phase transition occurs we observe a second peak at higher energy (1665 meV), corresponding to the orthorhombic phase (OP)\cite{Incenzo14, Fang15, Kong15}. It is noticeable, however, that the emission peak at lower energy ($\sim$1590 meV) remains dominant down to much lower temperatures than those where the edge can be seen in absorption (as shown in S.I. Fig. 1 (a)) and only disappears at temperatures below $\sim$ 80\,K.
The main peak also has a small shoulder on the low energy side at $\sim$1635 meV, which
suggests that emission from bound excitons, or possibly donor-acceptor-pair transitions\cite{Kong15}, may also be occurring in the majority, orthorhombic, phase. Emission from the low energy peak, which we infer is due to microcrystallite domains of the high temperature tetragonal phase (TP), is however found to have a strong spatial dependence.
Low temperature (4\,K) mapping results for MAPbI$_{3}$ are presented in Fig.~\ref{fig3} where the intensity of the blue colour in panel (a) indicates the integrated intensity of the dominant OP peak. The emission is relatively uniform while
still showing some variation on the scale of a few $\mu$m due to the underlying microcrystalline structure, as seen at high temperature. In addition, however,
the low energy, TP, peak at $\sim$1585 meV can still be detected at certain positions within
the sample. Typical spectra for two different points on the sample are shown in panel (c), where in one location we observe peaks from both phases and in the other only one is visible. The emission energy of the TP peak, approximately 80 meV lower in energy, is shifted down by the same amount as the change in band gap observed at the phase transition temperature as shown in Fig. 1(b) of the SI. The intensity of the TP peak is always weaker than the intensity of the OP peak, however the presence of the TP peak does appear to cause some decrease in the intensity of the OP peak. In
panel (b), we show the integrated intensity of the TP peak in red superimposed upon the integrated intensity of the OP peak. This shows a rather sparse distribution of isolated regions where the TP peak is present with a typical size of order 1-5 $\mu$m,
usually where the intensity of the OP peak is weaker. To test for correlation between the two
peaks, in panel (d) we plot the proportion of the total emission in each peak ($\frac{I_{OP,TP}}{(I_{OP}+I_{TP})}$) as a function of the emission intensity of the TP peak $(I_{TP})$. As the proportion of the emission in the TP peak (orange points) rises, the OP phase falls progressively.

Overall this behaviour provides strong evidence for the coexistence of the two crystal phases at low temperature, with the existence of a small number of anti-phase microdomains of the high temperature tetragonal phase which are capable of preferentially collecting free carriers due to their smaller band gap as suggested previously for MAPbI$_{3-x}$Cl$_{x}$ \cite{Wehrenfennig14b}. This is consistent with the observation of a dominant TP peak in PL at temperatures significantly below the phase transition, whereas it cannot be observed directly in absorption. Typical separations of the tetragonal inclusions are of the order of $10-20 \mu \textrm{m}$, suggesting that the diffusion length in this material remains at around this value until the temperature falls below 80\,K and that inter grain diffusion is significant in this temperature regime. Since the $\mu$PL is obtained by scanning the exciting laser and the diffusion length will be $\propto\sqrt{T}$ at low temperatures, the low temperature maps provide an upper limit to the size of high temperature phase domains of a few $\mu$m, which is comparable with the size of individual crystallites. In practice, they may be even smaller than this since the relatively long diffusion lengths observed in perovskites\cite{Xing13,Stranks13} mean that even microscopically small regions of smaller band gap material can be expected to collect excitons over distances comparable to the diffusion length. As a consequence these small regions can dominate the PL response without being detectable in absorption measurements.  The overall proportion of the high temperature tetragonal phase can thus be estimated to be no more than a few per cent of the total area and possibly much less. Such co-existing crystal phases have been observed previously in a number of different perovskite materials, often associated with the presence of strain\cite{Zeches09, Chen10}.

\begin{figure}[h]
\includegraphics[width=15cm]{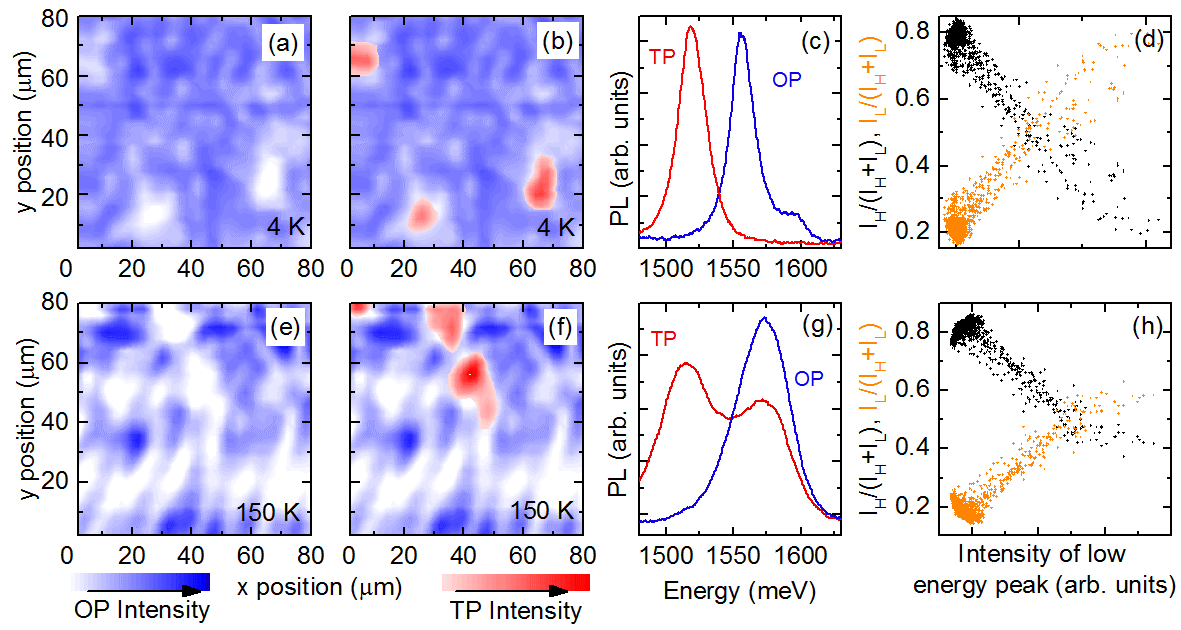}
\caption{\label{fig4} \textbf{Spatially resolved
micro-photoluminescence for MAPbI$_{3-x}$Cl$_{x}$}:(a)Integrated
intensity of the high energy (OP) peak as a function of position. The
intensity increases from white to dark blue.  (b) A second map of the low energy (TP) peak is superimposed onto the map shown in (a). The intensity increases from white to red. (c) Typical spectra for positions where only the OP peak was observed (blue) and where the TP peak was seen also (red). (d) Correlation of the integrated
intensity of the TP (orange) and OP (black) peaks versus
intensity of the low energy TP peak. (a)-(d) were performed at 4K, spot illumination intensity 0.3 nW, and (e)-(h) show the same measurements performed at 150K, spot illumination intensity 3 nW. The scanning step was
$2 \mu$m. }
\end{figure}

The presence of high temperature, tetragonal phase, domains becomes more significant in MAPbI$_{3-x}$Cl$_{x}$ as shown in Fig.~\ref{fig4} where the spatial mapping measured at 4\,K and 150\,K is shown.  In both cases the spectra show at least two well resolved emission peaks
at different spatial positions as shown in panels (c) and (g).
Panels (a) and (e) show maps at 4\,K and 150\,K respectively, based on the intensity of the OP peak, which
as above corresponds to the low temperature crystal phase. We observe the same the grain structure as at room temperature with typical grain sizes in the region of 5-10 $\mu$m. However, when we superimpose an additional
intensity map for the TP peak as shown in the panels (b)
and (f) for 4\,K and 150\,K respectively, we observe that the
areas with weak emission from the low temperature phase correspond very clearly with strong emission from the TP peaks.  This is particularly obvious at 4\,K, where almost the whole sample becomes
uniformly emissive from either one peak or the other. This suggests a
strong anti-correlation between the two sets of domains which we associate with individual crystallites being in either one crystal phase or the other. The anti
correlation is confirmed by plotting ($\frac{I_{OP,TP}}{(I_{OP}+I_{TP})}$) as before in Fig.~\ref{fig4}(d) and (h) for 4\,K and 150\,K respectively. In both cases there is a clear anti correlation, when the intensity of the TP peak reaches its maximum, the intensity of the OP peak is minimal. The larger scale of the crystallite domains also allows the two different phases to be completely resolved. Fig.~\ref{fig1}(b) and (c) show the completely different temperature dependence of the emission spectra for two microdomains which in Fig.~\ref{fig1}(b) transforms from the tetragonal to the orthorhombic phase at low temperature, while the region in Fig.~\ref{fig1}(c) remains in the tetragonal phase for the whole temperature range. In Fig.~\ref{fig1}(b), all sign of the TP peak disappears below 120\,K and the low temperature, orthorhombic phase initally shows a higher energy peak at around 1600 meV which we attribute to the free exciton, which is replaced below 60\,K by a lower energy peak at $\sim$1560 meV, presumed to be some form of bound exciton state.  By contrast Fig.~\ref{fig1}(c) shows spectra where the strongest feature is always the TP peak which becomes progressively more dominant as the temperature falls. By 30\,K only a single peak can be seen at $\sim$1525 meV which is again shifted down by the energy difference between the orthorhomic and tetragonal phases at the high temperature and no trace of the OP peak can be detected. Comparing the relative areas shown by the red and blue regions in Fig.~\ref{fig4}(f) allows us to estimate that similarly to the MAPbI$_{3}$, a few percent of the total sample remains in the tetragonal phase at low temperatures, but both the scale of the domains and their separation has increased by a factor consistent with the increased size of the microcrystallites.

In addition to the apparently random presence of the anti-phase domains shown in Figures \ref{fig3} and \ref{fig4}, there is also evidence that the presence of the different phases can be significantly influenced by local perturbations to the layer morphology.  This is shown in Fig. 2 and Fig. 3 of the S.I. for MAPbI$_{3-x}$Cl$_{x}$ where the low temperature image Fig. 2 (S.I.) shows a very clear physical pattern, associated probably with a crack or fissure of over 100 $\mu$m in length. This is visible as a positive image for the tetragonal phase emission at 1535 meV and as an equivalent negative image from the intensity map for the free and bound exciton peaks at 1565 and 1600 meV from the dominant orthorhombic low temperature phase. Fig. 3 (S.I.) shows a further low temperature image, where a small region of material has been photo-annealed with a focussed spot of 532 nm radiation creating local damage or phase modification.  At the centre of the annealing there is a dark core around which the emission becomes dominated by a domain of the tetragonal high temperature phase, followed sequentially by emission rings from the OP bound exciton, the OP free exciton and finally back to a macroscopically uniform region dominated by the OP bound exciton again, as shown previously in Fig \ref{fig1}.  It has been observed \cite{deQuilettes16, Hoke15} that photo-annealing causes iodine diffusion leading to a reduction in iodine density in the centre of a focussed spot surrounded by an increase in iodine content.  This would suggest that the local high temperature phase inclusions are strongly associated with a reduced iodine content, possibly associated with an enhanced level of iodine vacancies which are known to influence charge trapping and nonradiative recombination \cite{Du15} and that the iodine content is also responsible for influencing strongly the relative emission strength from bound and free excitons in the low temperature phase. This would also be consistent with the idea that the iodine content could be significantly perturbed close to the edges of cracks formed during film deposition.

We have also repeated the $\mu$PL studies for FAPbI$_{3}$.  This shows a comparable behaviour at low temperatures (S.I. Fig. 4(a)) to that seen for MAPbI$_{3}$ with a few small microcrystallite grains on a scale of 1-5 $\mu$m, which remain within the high temperature, tetragonal, phase giving a TP peak emission, and occupying at most a few percent of the total sample area. It also demonstrates a similar role played by the presence of 'cracks' as can be seen in Fig.~SI 4(b) at 4\,K.  There is a large dark crack or scar running through the centre of the image which is bordered by a region of tetragonal material, again suggestive of an enhanced concentration of iodine vacancies. Beyond this the material returns to the same distribution of a few microcrystallite grains of high temperature phase in the main body of the sample.

\begin{figure}[h]
\includegraphics[width=15cm]{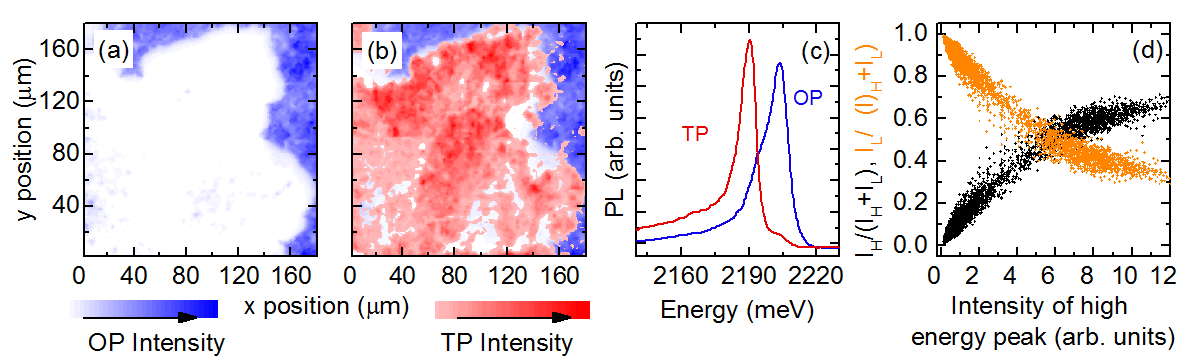}
\caption{\label{fig5} \textbf{Spatially resolved
micro-photoluminescence for FAPbBr$_{3}$}:(a)Integrated intensity of
the high energy (OP) peak as a function of position. The intensity increases
from white to dark blue. The scanning step was $2 \mu$m. (b) A second map of the low energy (TP) peak is superimposed onto the map shown in (a). The intensity increases from white to red. (c) Typical spectra for positions where
only the OP peak was observed (blue) and where both peaks were observed
(red). (d) Correlation of the integrated intensity of the TP
(orange) and OP (black) peaks versus intensity of the TP peak. All the measurements were performed at 4K, spot illumination intensity 5 nW.}
\end{figure}

We now turn to the case of FAPbBr$_{3}$, which shows significantly different behaviour to that seen for the iodides and the MAI$_{3-x}$Cl$_{x}$ alloy. The temperature dependence of the emission spectra for one region of the sample is shown in Fig.~\ref{fig1}(d). In this case there is no clear separation of two peaks which can be easily associated with the different phases but the spectra do show the disappearance of a low energy component at around the phase transition temperature of 150 - 160\,K, leaving two strong peaks at around 2190 and 2205 meV at low temperature.  The room temperature $\mu$PL mapping as shown in Fig.~\ref{fig2} demonstrates that there is a similar grain structure and size to that seen for the MAPbI$_{3}$ and FAPbI$_{3}$. The low temperature mapping shown in Fig.~\ref{fig5}, demonstrates, however, completely different behaviour.  In this case we see variations in the emission spectra on a much larger physical scale as compared to that seen for the iodide samples. This suggests that the phase domain structure for the bromide is independent of the grain structure of the layer. Fig.~\ref{fig5}(a) shows that when imaged at 2205 meV, large areas of the film do not emit. Imaging at 2190 meV, however, we see in Fig.~\ref{fig5}(b) that there is a complementary emission, as marked in red, which corresponds very well with the non emitting area of the low temperature orthogonal phase and Fig.~\ref{fig5}(c) shows that the spectra from the two different regions are well resolved. This suggests that macroscopic domains exist consisting of multiple microcrystallites, which can be either orthorhombic or tetragonal at low temperatures with typical domain sizes on the order of 100 $\mu$m or more.  The difference in the emission energies of only 15 meV between the two regions suggests that there is very little difference in band structure or free energy between the two phases as has been shown previously\cite{Kunugita15}, which would allow the phase transition to propagate more easily thus creating macroscopic domains.

In conclusion we have demonstrated the existence of a microscopic phase domain structure in the tri-iodide organic-inorganic perovskites at temperatures well below the tetragonal to orthorhombic phase transition. The scale of the domains is strongly related to the underlying microcrystalline structure and corresponds to individual crystallites which adopt one or the other crystal structure, such that a few percent of the final layer remains within the high temperature tetragonal phase, even at 4\,K. The existence of even a small proportion of material in these domains can allow them to dominate the emission properties, particularly at higher temperatures, where the diffusion lengths can be very long\cite{Xing13,Stranks13}. This is likely to be very significant for optoelectronic devices such as perovskite based lasers\cite{Saliba15}.  By contrast the tri-bromide FAPbBr$_{3}$ shows a much more macroscopic phase domain structure which is independent of the crystallites, probably associated with the very small energetic difference between the two crystal structures, which suggests that the tri-Bromides might be significantly better in device applications.  We have also demonstrated that the presence of strains or local damage can have a significant effect on the domain structure, producing localised regions of phase modification, which also suggests that elimination of such features should be an important goal for device optimisation.

\section{Experimental Method}

Perovskite precursor synthesis: Formamidinium iodide (FAI) and formamidinium bromide (FABr) were synthesised by
dissolving formamidinium acetate powder in a 1.5$x$ molar excess of 57$\%$ w/w hydroiodic acid (for FAI) or 48$\%$ w/w
hydrobromic acid (for FABr). After addition of acid the solution was left stirring for 10 minutes at 50$^{\circ}$ C.  Upon
drying at 100$^{\circ}$C, a yellow-white powder is formed. This was then washed twice with diethyl ether and recrystallized
with ethanol, to form white needle-like crystals. Before use, it was dried overnight in a vacuum oven. To form
FAPbI$_{3}$ and FAPbBr$_{3}$ precursor solutions, FAI and PbI$_{2}$ or FABr and PbBr$_{2}$ were dissolved in anhydrous
N,N-dimethylformamide (DMF) in a 1:1 molar ratio, at 0.55M of each reagent, to give a 0.55M perovskite precursor
solution. To form CH$_{3}$NH$_{3}$PbI$_{3-x}$Cl$_{x}$ precursor solutions, Methylammonium iodide (MAI) and lead
chloride (PbCl$_2$) were dissolved in a 40$\%$ w/w DMF solution in a 3:1 molar ratio. For MAPbI$_{3}$, precursor
solutions were prepared separately by dissolving lead iodide (PbI$_2$) in DMF (450 mg/ml), and MAI in isopropanol (50
mg/ml), respectively.

Film formation: All the samples were prepared in a nitrogen-filled glovebox on glass substrates cleaned sequentially in
hallmanex, acetone, isopropanol and O$_{2}$ plasma. Immediately prior to the Formamidinium (FA) sample film formation,
small amounts of acid were added to the precursor solutions to enhance the solubility of the precursors and allow
smooth and uniform film formation. 38$\mu$l of hydroiodic acid (57$\%$ w/w) was added to 1ml of the 0.55M FAPbI$_{3}$
precursor solution, and 32$\mu$l of hydrobromic acid (48$\%$ w/w) was added to 1ml of the 0.55M FAPbBr$_{3}$ precursor
solution. FA Films were then spin-coated from the precursor plus acid solution on warm (85$^{\circ}$C) substrates for 45s
at 2000rpm, followed by annealing at 170$^{\circ}$C in air for 10 minutes. This gave very uniform pinhole-free layers,
~350nm thick, of FAPbI$_{3}$ or FAPbBr$_{3}$.

The Methylammonium (MA) samples were prepared following methods described previously \cite{Lee12, Heo13}. In brief,
CH$_{3}$NH$_{3}$PbI$_{3-x}$Cl$_{x}$ films were prepared by spin-coating for 60s at 2000 rpm on warm (85$^{\circ}$C)
substrates, followed by annealing at 100$^{\circ}$C in air for 1 hour. The CH$_{3}$NH$_{3}$PbI$_{3}$ films were prepared
from a PbI$_2$ layer which was first deposited on cleaned glass by spin-coating at 6000rpm for 30s from a precursor
solution, followed by drying at 70$^{\circ}$C for 5 min. Then the MAI layer was deposited on the dried PbI2 layer by
spin-coating at 6000rpm for 30s from a precursor solution, followed by annealing at 100$^{\circ}$C for 1 hour.

All films were all sealed by spin-coating a layer
of the insulating polymer poly(methyl methacrylate) (PMMA) at
1000rpm for = 60s (precursor solution 10mg/ml in cholorobenzene) on
top in order to ensure air-and moisture-insensitivity.

For the optical measurements the sample was placed in a helium flow cryostat with optical access with the accessible
temperature range $3-300$ K. Excitation and collection was implemented using a microscope objective with a numerical
aperture $NA = 0.66$ and magnification $50\times$. The typical diameter of the spot was of the order of $1 \mu$m.
Additionally, the cryostat was mounted on motorized $x-y$ translation stages to allow high resolution spatial mapping.
The $\mu$PL spectra have been recorded using a spectrometer equipped with a CCD camera. A green solid-state laser,
emitting at $532$ nm, was used for excitation.

\vspace{0.5cm}

\noindent\textbf{Acknowledgments}

\noindent  This work was partially supported by the Region
Midi-Pyren\'ees, the Programme Investissements d'Avenir under the contract MESR, 13053031, project BLAPHENE under IDEX program Emergence,  ANR JCJC project milliPICS, EuroMagNET II under EU Contract
228043, Meso-superstructured Hybrid Solar
Cells -MESO NMP-2013-SMALL7-604032 project, the Engineering and
Physical Sciences Research Council (EPSRC) and the European Research
Council (ERC-StG 2011 HYPER Project no. 279881).

\vspace{0.5cm}

\noindent\textbf{Author contributions}

\noindent The samples were prepared by J.T-W.W. and G.E.E. The measurements have
been performed by K.G, A.M, A.S, Z.Y. and P.P.  The first draft of the
paper was written by P.P and R.J.N. with all authors contributing to the
final version.

\vspace{0.5cm}

\noindent\textbf{Competing Financial Interests}

\noindent The authors declare that they have no competing financial interests.

\vspace{0.5cm}

\noindent\textbf{References}

%

\clearpage

\textbf{Figures}

\clearpage

\section{Supplementary Information: Spatially resolved studies of the
phases and morphology of methylammonium and formamidinium lead
tri-halide perovskites}

\author{K. \surname{Galkowski}}
\affiliation{Laboratoire National des Champs Magn\'etiques Intenses,
CNRS-UJF-UPS-INSA, 143, avenue de Rangueil, 31400
Toulouse}\affiliation{Institute of Experimental Physics, Faculty of
Physics, University of Warsaw - Pasteura 5, 02-093 Warsaw, Poland}

\author{A.  \surname{Mitioglu}}
\affiliation{Laboratoire National des Champs Magn\'etiques Intenses,
CNRS-UJF-UPS-INSA, 143, avenue de Rangueil, 31400 Toulouse}
\affiliation{Institute of Applied Physics, Academiei Str. 5,
Chisinau, MD-2028, Republic of Moldova}

\author{A.  \surname{Surrente}}
\affiliation{Laboratoire National des Champs Magn\'etiques Intenses,
CNRS-UJF-UPS-INSA, 143, avenue de Rangueil, 31400 Toulouse}

\author{Z.  \surname{Yang}}
\affiliation{Laboratoire National des Champs Magn\'etiques Intenses,
CNRS-UJF-UPS-INSA, 143, avenue de Rangueil, 31400 Toulouse}

\author{D. K. \surname{Maude}}
 \affiliation{Laboratoire National des Champs Magn\'etiques Intenses,
CNRS-UJF-UPS-INSA, 143, avenue de Rangueil, 31400 Toulouse}

\author{P. \surname{Kossacki}}
\affiliation{Institute of Experimental Physics, Faculty of Physics,
University of Warsaw - Pasteura 5, 02-093 Warsaw, Poland}

\author{G. E. \surname{Eperon}}
 \affiliation{University of Oxford, Clarendon
Laboratory, Parks Road, Oxford, OX1 3PU, United Kingdom}

\author{J. T-W.\surname{Wang}}
 \affiliation{University of Oxford, Clarendon
Laboratory, Parks Road, Oxford, OX1 3PU, United Kingdom}

\author{H. J. \surname{Snaith}}
\affiliation{University of Oxford, Clarendon Laboratory, Parks Road,
Oxford, OX1 3PU, United Kingdom}

\author{P. \surname{Plochocka}}
\email{paulina.plochocka@lncmi.cnrs.fr}\affiliation{Laboratoire
National des Champs Magn\'etiques Intenses, CNRS-UJF-UPS-INSA, 143,
avenue de Rangueil, 31400 Toulouse}

\author{R. J. \surname{Nicholas}}
\email{robin.nicholas@physics.ox.ac.uk}\affiliation{University of
Oxford, Clarendon Laboratory, Parks Road, Oxford, OX1 3PU, United
Kingdom}


\maketitle

Fig. \ref{Fig1SI} of the supplementary information presents
macro-transmision (transmission of white light averaged over 1
mm$^2$ spot)  measured as a function of the temperature for
MAPbI$_{3}$. Panel (a) contains the transmission spectra taken at
different temperatures, and panel (b) shows resulting absorbtion
energies plotted as a function of the temperature. At temperatures
around the phase transiton  (150-170 K) the spectra reveal presence
of both orthorhombic (orthorhombic phase, OP) and tetragonal
(tetragonal phase, TP) phases. The phase transition from tetragonal
to orthorombic phase causes an abrupt rise of around 100 meV in the
energy of the absorbtion edge.  The temperature dependence of the TP
peak reveals a turning point at 168 K (b). No trace of the OP is
found above 170 K in transmission and in both temperature dependent
and spatially resolved $\mu$-PL studies, suggesting that above 168 K
the material is entirely in the tetragonal phase.

\begin{figure}[h]
\includegraphics[width=14cm]{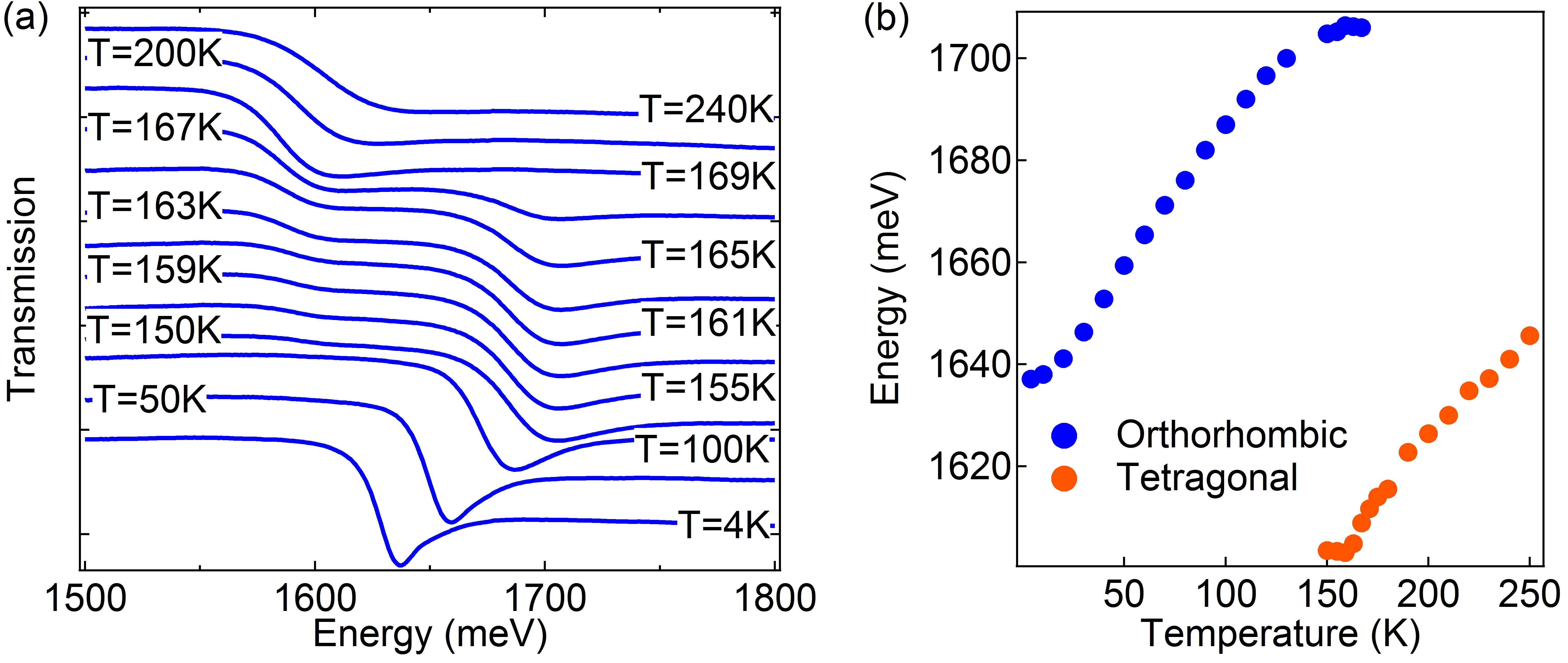}
\caption{\label{Fig1SI} \textbf{Temperature dependence for
MAPbI$_{3}$ in macro transmission}. (a) Transmission spectra and (b)
energy of the absorbtion edge as a function of temperature. }
\end{figure}

\begin{figure}[h]
\includegraphics[width=14cm]{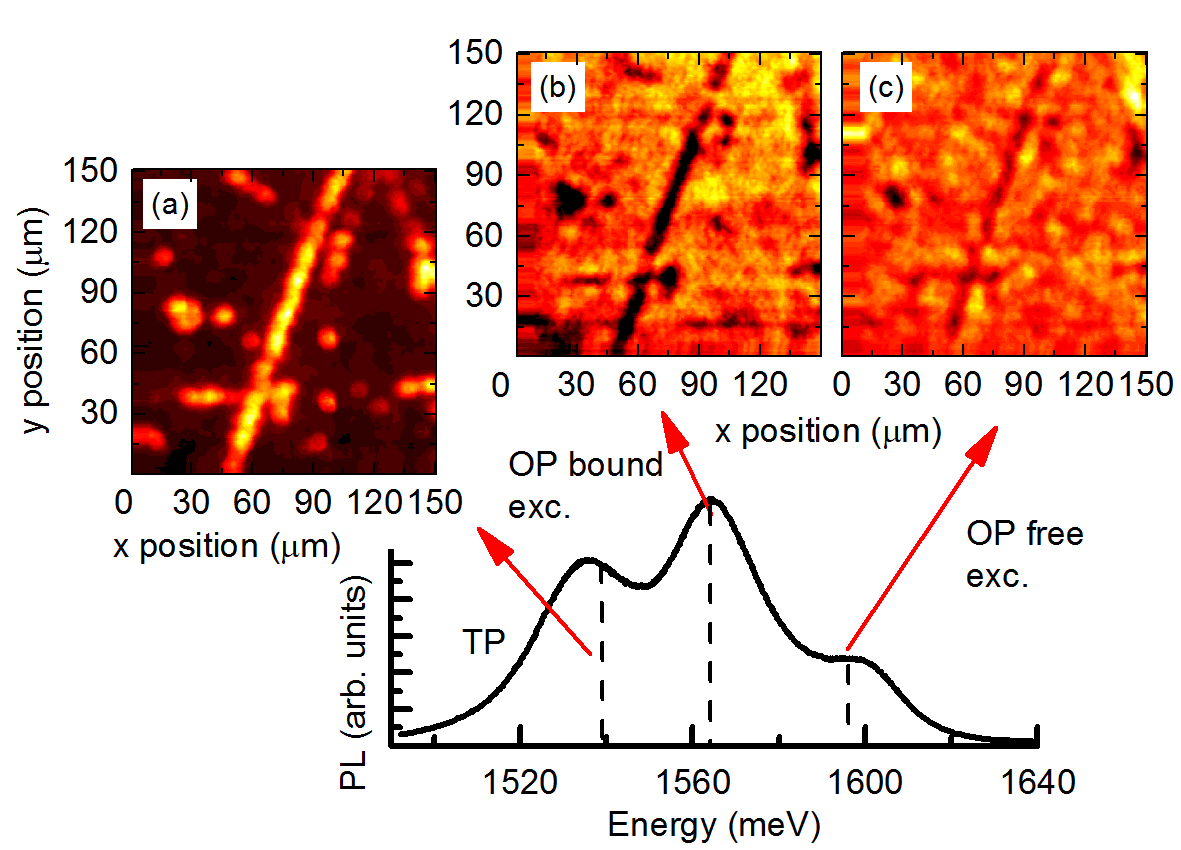}
\caption{\label{Fig2SI} \textbf{Spatially resolved
micro-photoluminescence for MAPbI$_{3-x}$Cl$_{x}$ at $T=5$ K,
scanning step 1 $\mu$m }. Photoluminescence spectrum and integrated
intensity maps of the sample corresponding to: (a) high temperature,
TP peak and bound (b) and free (c) exciton peaks of the low
temperature OP.}
\end{figure}

Fig. \ref{Fig2SI} shows a low-temperature (5 K) $\mu$-PL map of a
selected region of a MAPbCl$_{x}$I$_{3-x}$ sample. This demonstrates
an obvious positive/negative correlation between the peaks at 1535
meV and 1565/1600 meV. As stated in the main article, we attribute
the 1535 meV emission (panel a) to high temperature, tetragonal
phase and the 1565 (panel b)/1600 meV(panel c) peaks respectively to
bound and free exciton of the low temperature, orthorhombic phase.
The positive/negative correlation of the OP and TP peaks is
particularly clear for the narrow-strip regions with well-defined
borders (compare panels (a) and (b)). This may suggest an enhanced
presence of the of high temperature phase in strained or cracked
areas.  There is also significantly more variation in the intensity
of the bound exciton when compared to the free exciton for the OP
phase, which is found to be most strongly quenched in the regions
close to the tetragonal phase inclusions (c).

\begin{figure}[h]
\includegraphics[width=14cm]{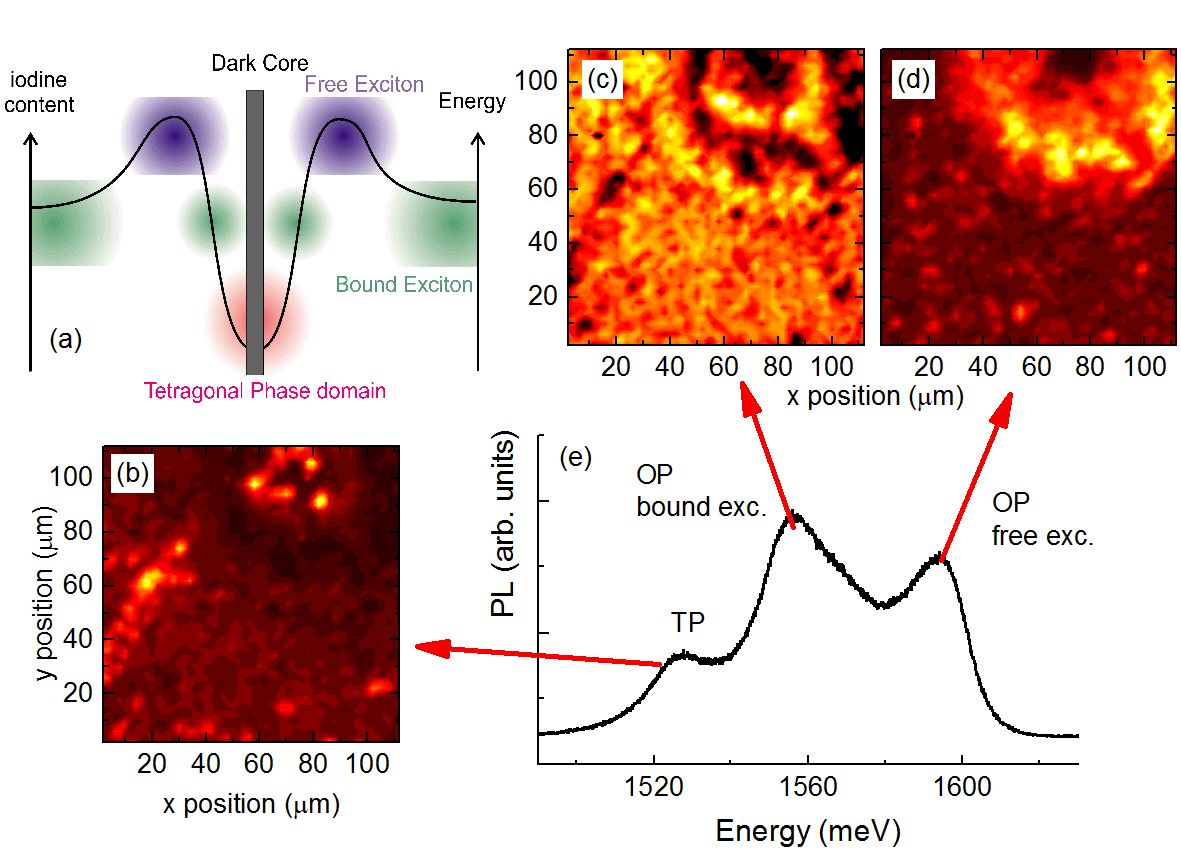}
\caption{\label{Fig3SI} \textbf{Spatially resolved
micro-photoluminescence for MAPbI$_{3-x}$Cl$_{x}$  at $T=10$ K,
scanning step 2 $\mu$m } (a) the distribution of iodine ions as a
function of distance from the point of exposure. (b)-(d)integrated
intensity maps of the sample corresponding to: (b) high temperature,
TP peak and bound (c) and free (d) exciton peaks of the low
temperature OP. (e) Photoluminescence spectrum showing all observed
transitions.}
\end{figure}

Further conclusions about the nature of observed inhomogenities come
from an investigation of regions exposed to a focused laser spot. It
has been reported recently \cite{deQuilettes16}, that photo -
annealing induces the migration of halide ions away from the point
of exposure, influencing the PL properties of the material. Fig.
\ref{Fig3SI}  presents low temperature spatially resolved
micro-photoluminescence for a MAPbCl$_{x}$I$_{3-x}$ sample
illuminated for several minutes with a 532 nm laser spot of  $10^9$
W/m$^{2}$ power density at room temperature. As a result of photo
annealing, regions of decreased and enhanced halide content are
formed Fig. \ref{Fig3SI} (a), leading to emissions characteristic of
both the tetragonal phase and both excitons of the orthorhombic
phase (e). The center of the laser spot is a dark core (upper right
corner of the maps). Moving away from the dark core (with a probably
increasing iodine content), we find regions of TP emission (b).
Sequentially, we observe rings of emission from OP bound excitons,
OP free excitons (d) and then again OP bound excitons which are
dominant in the unaffected areas. This picture is consistent with
the emission being controlled by an iodine content as shown in panel
(a), where OP free excitons are only dominant at low temperatures in
regions of enhanced iodine content.

\begin{figure}[h]
\includegraphics[width=14cm]{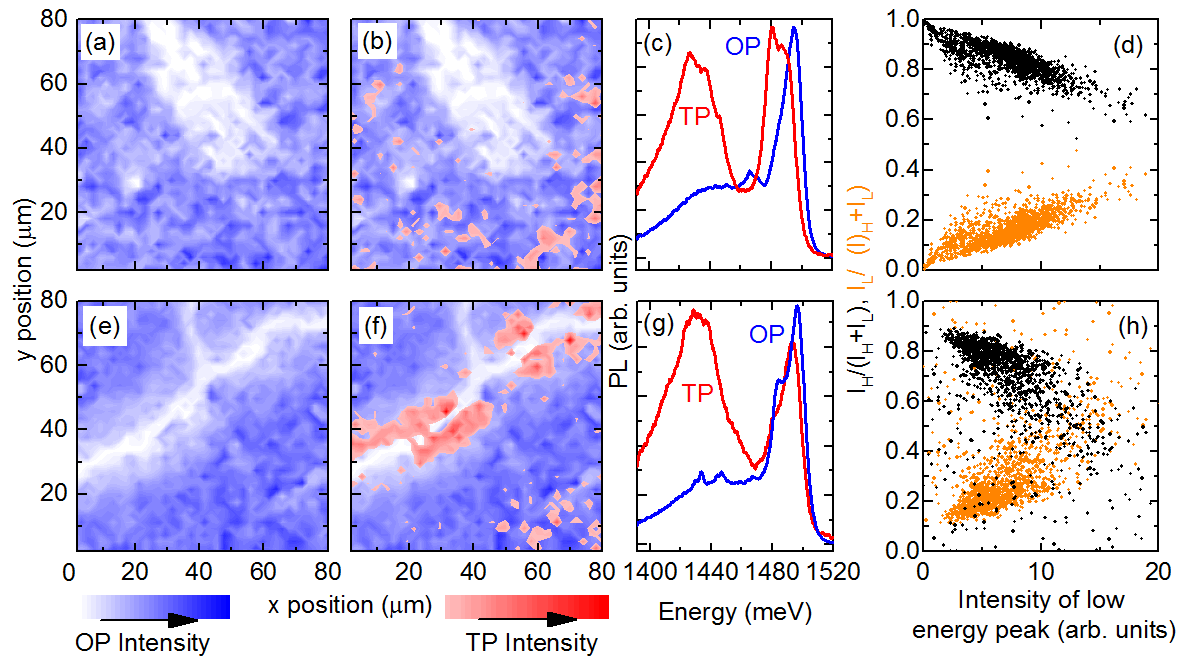}
\caption{\label{Fig4SI} \textbf{Spatially resolved
micro-photoluminescence for FAPbI$_{3}$ at $T=5$ K, scanning step $2
\mu$m.}. The images present a homogenous area (a)-(d) and a regon
around dark ribbon (e)-(h).  (a),(e): Integrated intensity of the OP
peak across the sample. The intensity increases from white to dark
blue. (b),(f): Two maps of the integrated intensity of the OP and TP
peaks. The intensity of the OP (TP) is marked in blue (red)
respectively. (c),(g): Typical photoluminescence spectra with
dominance of OP (blue) and TP (red) peaks. (d),(h): Correlation of
the integrated intensity of the TP (orange) and OP (black) peaks
versus intensity of the TP peak.}
\end{figure}

We supplement the spatially resolved  $\mu$-PL obtained for
MAPbI$_{3}$ with measurements performed on compounds based on the
Formamidinium cation, FAPbI$_{3}$ (Fig.\ref{Fig4SI}). We show a
predominantly uniform region (panels a- d) and an area in proximity
of a dark ribbon, being probably a crack or fissure (panels d-h).
Similarly to MAPbI$_{3}$ , at low temperatures we observe the anti
correlation of the OP and TP emission peaks, with TP phase grains up
to 10 $\mu$m in size (panels a and b). Both bound and free excitons
of OP occur simultaneously in the investigated region (c). The
regions close to the dark ribbobn (panels (d)-(e)) show a 1D analogy
of the photo-annealing results. Close to the ribbon, we observe
enhanced  emission from the TP peak, with additional narrow emission
lines probably from localized states (g). In addition the TP/OP
anticorrelation is significantly stronger than in the case of the
region shown in (a) (compare panels d and h). This gives further
evidence to link the existence of the tetragonal phase inclusions
with the presence of strained or defected areas.

\vspace{0.5cm}


%
%
%

\end{document}